\documentclass[12pt]{article}
\usepackage{epsfig}
\usepackage{amsfonts}
\usepackage{latexsym}
\usepackage{amsmath,amssymb}
\usepackage{mathrsfs}
\usepackage{hyperref}
\usepackage{setspace}
\usepackage{color}
\usepackage{bm}
\usepackage{slashed}
\usepackage{braket}
\numberwithin{equation}{section}

\newcommand{\x}{{\bf x}}
\newcommand{\A}{{\bf A}}
\newcommand{\J}{{\bf J}}

\newcommand{\tf}{t_{\rm f}}
\newcommand{\dl}{\Delta L}
\newcommand{\uv}{\Omega_{\rm m}}
\newcommand{\Mpl}{M_{\rm pl}}
\newcommand{\tdec}{\tau_{\rm dec}}
\newcommand{\Ts}{T_{\rm s}}
\newcommand{\Tc}{T_{\rm c}}

\begin{document}

\begin{titlepage}
\unitlength = 1mm
\begin{flushright}
OU-HET-1065\\
KOBE-COSMO-20-12
\end{flushright}

\vskip 1cm
\begin{center}

{\large {\textsc{\textbf{Noise and decoherence induced by gravitons}}}}

\vspace{1.8cm}
Sugumi Kanno$^*$,  Jiro Soda$^{\flat}$ and Junsei Tokuda$^{\flat}$

\vspace{1cm}

\shortstack[l]
{\it $^*$ Department of Physics, Osaka University, Toyonaka 560-0043, Japan \\ 
\it $^\flat$ Department of Physics, Kobe University, Kobe 657-8501, Japan
}

\vskip 1.5cm

\begin{abstract}
\baselineskip=6mm

We study quantum noise and decoherence induced by gravitons. We derive a Langevin equation of geodesic deviation in the presence of gravitons. The amplitude of noise correlations tells us that large squeezing is necessary to detect the noise. We also consider the decoherence of spatial superpositions of two massive particles caused by gravitons in the vacuum state and find that gravitons could give a relevant contribution to the decoherence. The decoherence induced by gravitons would offer new vistas to test quantum gravity in tabletop experiments.

\end{abstract}

\vspace{1.0cm}

\end{center}
\end{titlepage}

\pagestyle{plain}
\setcounter{page}{1}
\newcounter{bean}
\baselineskip18pt

\setcounter{tocdepth}{2}

\tableofcontents

\section{Introduction}

An understanding of the nature of gravity has been a central issue in physics
since the discovery of general relativity and quantum mechanics.
Nevertheless, no one has succeeded in constructing quantum theory of gravity. 
In particular, the existence of  gravitons is still obscure~\cite{DYSON:2013jra}.
In these situations, it is legitimate to doubt the necessity of canonical quantization
of gravity~\cite{Jacobson:1995ab}.  
Hence, it is worth seeking an experimental evidence of
quantum gravity. Usually, theorists explore the field of quantum gravity at energy scales near the Planck scale. However, it is far beyond the capacity of the current or future particle accelerators. 
Instead,  cosmological observations have been exploited  for probing the high-energy physics. In fact, cosmological observations suggest that the large scale structure of the universe stems from the quantum fluctuations during the inflationary stage. It is natural to consider that primordial gravitational waves are also generated directly from the quantum fluctuations. Hence, one possible approach to testing quantum gravity is to study the nonclassicality of primordial gravitational waves~\cite{Kanno:2018cuk,Kanno:2019gqw}. 
Recently, as an alternative approach, tabletop experiments are drawing attention~\cite{Carney:2018ofe,Ito:2019wcb,Ito:2020wxi}.
Remarkably, based on the development of quantum information, several ideas to test the quantum nature of gravity through laboratory experiments are proposed~\cite{Bose:2017nin,Marletto:2017kzi}. 
More recently, as a new probe of gravitons, noise in the lengths of the arms of gravitational wave detectors is discussed by using a path integral approach~\cite{Parikh:2020nrd}.
One of our goals in this paper is to derive the quantum Langevin equation in order  
to obtain the noise in the gravitational wave detectors.

The noise is usually associated with the decoherence induced by quantum entanglement between a system and gravitons~\cite{Calzetta:1993qe}. 
Thus, as an approach to testing quantum gravity, it would be important to understand the noise induced by gravitons and then the decoherence caused by the noise. 
 The decoherence due to gravity in the context of 
 quantum superposition of massive objects has been investigated~\cite{Bassi:2017szd}\!
 \footnote{The cosmological decoherence due to thermal gravitons can be found in~\cite{Bao:2019ghe}.}.
 The effect of gravitational field on the quantum dynamics of nonrelativistic particles was investigated by using the influence functional method and it is shown that the decoherence due to gravitational field is effective in the energy eigenstate basis~\cite{Anastopoulos:1995ya}. 
 Moreover, based on effective field theory approach, the decoherence rate was derived under the Markovian approximation (the assumption that the correlation time is very short)~\cite{Blencowe:2012mp}.
 The quantum Markov master equation for the gravitating matter was derived in \cite{Anastopoulos:2013zya}. 
  Following the paper of decoherence in the context of electromagnetic dynamics~\cite{Breuer and Petruccione},
  the effect of the gravitational bremsstrahlung on the destruction process of quantum superposition has been considered~\cite{Riedel:2013yca}. More explicit formulation along this direction has been given in \cite{Oniga:2015lro}. The formalism is further applied to the system of atoms~\cite{Oniga:2017pyq}.
 Recently, this possibility discussed again under the Markovian assumption~\cite{Vedral:2020dnh}.
 The decoherence due to quantum fluctuations of geometry caused by gravitons is also discussed in~\cite{DeLorenci:2014vwa}. Since the Markovian approximation cannot be applied to the decoherence caused by the gravitational bremsstrahlung, the non-Markovian decoherence process is analyzed in
 ~\cite{Suzuki:2015nva}.  In the above papers, the minimal coupling of a metric
 to a particle has been considered.
 However, from the point of view of the equivalence principle, the point particle does not feel gravity. Hence, the deviation of the geodesics is studied in~\cite{Quinones:2017wka}.
Thus, when considering the decoherence due to gravitons,  it would be necessary to take into account both of the non-Markov process and the equivalence principle.

In this paper, we study the quantum noise and decoherence in order to probe gravitons and ultimately quantum gravity. 
Firstly, we target on quantum noise in gravitational wave detectors. When gravitational waves arrive at the laser interferometer, the suspended mirrors interact with the gravitational waves. The mirror interacts with an environment of gravitons quantum mechanically. By using Fermi normal coordinates, we evaluate the effect of quantum noise induced by gravitons on the suspended mirrors. We show that the noise in the squeezed state can be sizable.
The results we obtain by using the quantum Langevin equation are consistent with those derived by using the path integral method in \cite{Parikh:2020nrd}.

Secondly, as our main goal in this paper, we consider a tabletop experiment by using two massive particles, one of which is superposed spatially, so called, the quantum state of Schr\"odinger's cat. Without using the Markovian assumption,
we give a formula for the decoherence rate of the superposition 
induced by gravitons. We then evaluate the decoherence rate for some simple configurations of superposition states and  
show that the decoherence due to gravitons could be a relevant contribution. To explore the decoherence process due to gravitons would be a first step toward discovery of gravitons in laboratory. 

The organization of the paper is as follows:
In section 2, we describe geodesics in the graviton background  
and derive a Langevin type equation of the system by eliminating
the environment of gravitons. 
In section 3, we evaluate the  noise correlation functions and show that
the noise can be observable if the gravitons are in the squeezed state.
In section 4, we discuss the decoherence induced by gravitons
and detectability of gravitons.
We give a formula for the decoherence rate and evaluate it for
simple cases.
The final section is devoted to the conclusion. 
A detailed calculation of a momentum integral is presented in the Appendix A and the derivation of decoherence functionals is given in Appendix B.
We work in the natural unit: $c=\hbar=1$.

\section{Quantum mechanics in the  graviton background}
\label{section2}

In this section, we present a model to study quantum mechanics 
in the graviton background. It gives rise to the basis for
studying the noise and  the decoherence  due to low energy gravitons. 
In particular, we derive the quantum Langevin equation.

\subsection{Gravitational waves}
\label{section2.1}

We consider  gravitational waves in the Minkowski space.
The metric describing gravitational waves in the transverse traceless gauge
 is expressed as
\begin{eqnarray}
ds^2=-dt^2+(\delta_{ij}+h_{ij}) dx^idx^j\,\,,
\label{metric}
\end{eqnarray}
where $t$ is the time, $x^i$ are spatial coordinates, $\delta_{ij}$ and $h_{ij}$ are the Kronecker delta and the metric perturbations which satisfy
the transvers traceless conditions $h_{ij,\,j}=h_{ii}=0$. The indices $(i,j)$ run from $1$ to $3$.  
Substituting the metric Eq.~(\ref{metric}) into the Einstein-Hilbert action, we obtain the quadratic action
\begin{eqnarray}
 \frac{1}{2\kappa^2}\int{\rm d}^4x \sqrt{-g}\,R
\simeq \frac{1}{8\kappa^2}\int{\rm d}^4x\,
\left[\,\dot{h}^{ij} \dot{h}_{ij}   -      h^{ij,k}\,     h_{ij,k}
\,\right]  \label{G_action}\, ,
\end{eqnarray}
where $\kappa^2=8\pi G$ and a dot denotes the derivative with respect to the time.
We can expand the metric field $h_{ij}( x^i, t)$ in terms of the Fourier modes
\begin{eqnarray}
h_{ij}(x^i ,t ) = \frac{2\kappa}{\sqrt{V}}\sum_{{\bf k}, A} \ h^A_{\bf{k}}(t)\,e^{i {\bf k} \cdot {\bf x}} \ e_{ij}^A(\bf k)  \,,
\label{fourier}
\end{eqnarray}
where we introduced the polarization tensor $e^A_{ij}({\bf k})$ normalized as $e^{*A}_{ij}({\bf k}) e^B_{ij} ({\bf k})= \delta^{AB}$. 
Here, the index $A$ denotes the linear polarization modes $A=+,\times$. 
Note that  we consider finite volume $V=L_{x }L_{y}L_{z}$ and 
 discretize the ${\bf k}$-mode with a width ${\bf k} = \left(2\pi  n_x/L_x\,,2\pi  n_y/L_y\,,2\pi  n_z/L_z\right)$ where ${\bf n}=(n_x , n_y , n_z)$ 
are integers.  
Substituting the formula \eqref{fourier} into the quadratic action \eqref{G_action}, we get
\begin{eqnarray}
S_g \simeq  \int dt \sum_{{\bf k}, A}  \left[\,\frac{1}{2} \dot{h}^A ({\bf{k}},t)  \dot{h}^A({\bf{k}},t)
 -   \frac{1}{2} k^2 h^A({\bf{k}},t) h^A({\bf{k}},t)  \,\right] \ ,
 \label{action-gravitons}
\end{eqnarray} 
where we used $k=|\bf k|$. 
 
\subsection{Action for two test particles}
\label{section2.2}

\begin{figure}[t]
\vspace{-1cm}
\includegraphics[height=7cm]{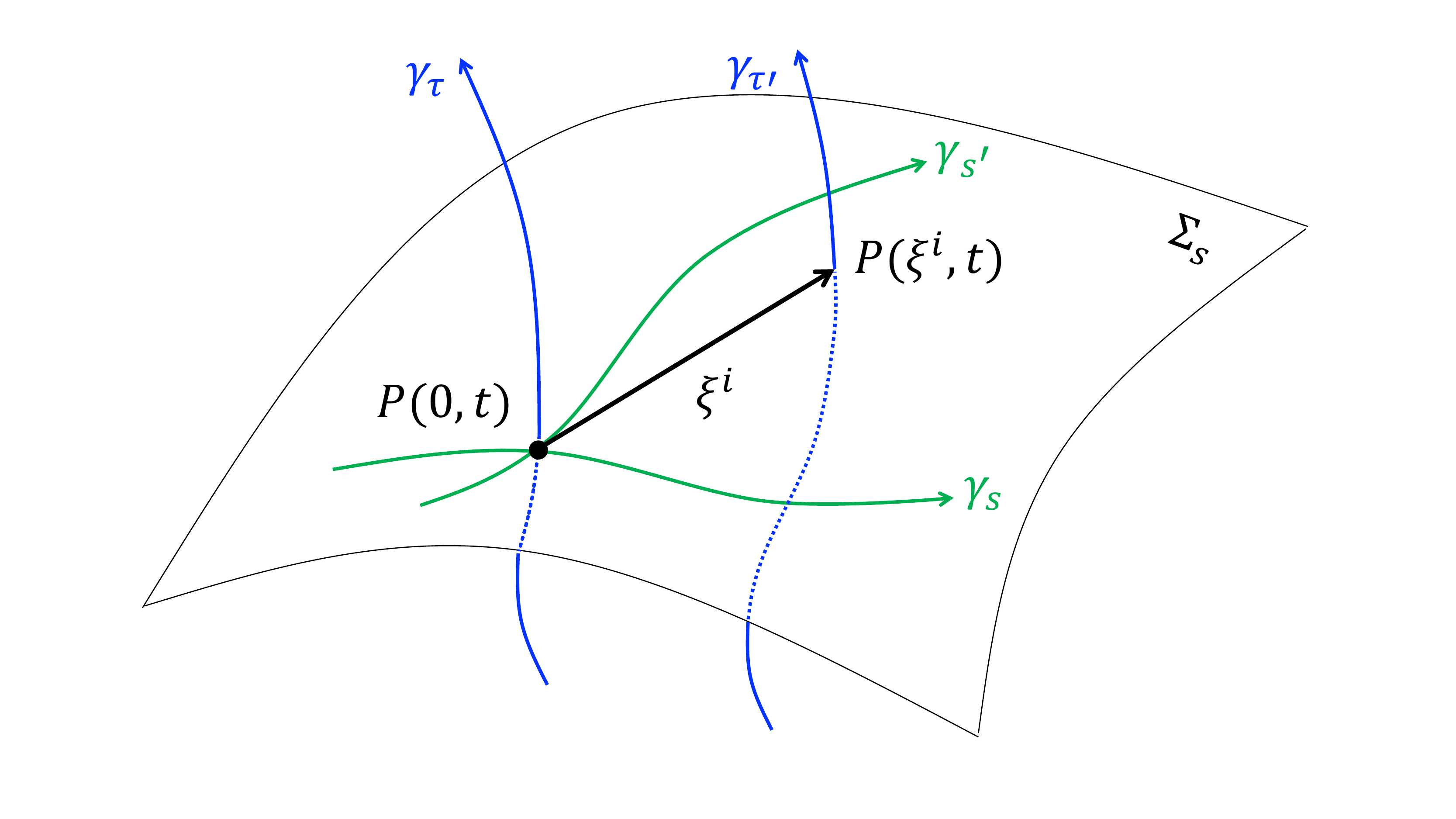}\centering
\vspace{-0.5cm}
\caption{Two neighboring timelike geodesics ($\gamma_\tau, \gamma_{\tau'}$) separated by $\xi^i$ are depicted in the blue lines and the green lines show two neighboring spacelike geodesics  ($\gamma_s, \gamma_{s'}$) orthogonal to the geodesic $\gamma_\tau$ in the spacelike hypersurface  $\Sigma_s$. We introduce the Fermi normal coordinate system using the orthogonal geodesics 
 at the point $P(0,t)$.} 
\label{fig1}
\end{figure}

When gravitational waves arrive at the laser interferometers, the suspended mirrors interact with the gravitational waves. Let us regard the mirror as a point particle for simplicity. A single particle, however, does not feel the gravitational waves because of the Einstein's equivalence principle at least classically. To see the effect of the gravitational waves, we need to consider two massive particles and measure the geodesic deviation between them. 

In this subsection, we evaluate the effect of gravitational waves on the two particles by introducing an appropriate coordinate system called the Fermi normal coordinates along one of their geodesics $\gamma_\tau$ (See Figure 1). The Fermi normal coordinate system represents a local inertial frame. The dynamics of the other geodesic of particle $\gamma_{\tau'}$ is described by the position $x^i(t)=\xi^i(t)$ in the vicinity of the point $P(0,t)$ 
and $\xi^i$ represents the deviation.

The action for the two test particles along the geodesics $\gamma_{\tau}, \gamma_{\tau'}$ is given by
\begin{eqnarray}
  S_p 
    =  - m\int_{\gamma_{\tau'}}d\tau 
    = -m \int_{\gamma_{\tau'}} dt \sqrt{-g_{\mu\nu}\left(\xi^i,t\right)\,\dot{\xi}^\mu\,\dot{\xi}^\nu} \,.
 \label{action1}
\end{eqnarray}
where $\xi^\mu=(t,\xi^i(t))$. Note that we omit the action for the particle along $\gamma_{\tau}$ because it's in the inertial frame and then the action has no dynamical variables.  The metric $g_{\mu\nu}$ up to the second order of $x^i$ in the Fermi coordinates is computed as
\begin{eqnarray}
ds^2 \simeq \left(- 1- R_{0i0j}\,x^i x^j \right) dt^2 -\frac{4}{3} R_{0jik}\,x^j x^k\, dt dx^i 
+\left(\delta_{ij} -\frac{1}{3}  R_{ikj\ell}\,x^k x^\ell \right) dx^i dx^j \ .
\label{metric2}
\end{eqnarray}
Here the Riemann tensor is evaluated at the origin $x^i =0$ in the Fermi normal coordinate system. Substituting the metric~(\ref{metric2}) into the action~(\ref{action1}), the action for the two particles up to the second order of ${\xi^i}$ is expressed as
\begin{eqnarray}
S_p  \simeq \int_{\gamma_{\tau'}} dt \left[ \frac{m}{2}  \dot{\xi}^{i^2} -  \frac{m}{2} R_{0i0j} (0,t)\,\xi^i \xi^j\right] \,.
\end{eqnarray}
Because the Riemann tensor $R_{0i0j}$ is gauge invariant at the leading order in the metric fluctuation $h_{ij}$, we can evaluate it in the transverse traceless gauge to get $R_{0i0j}(0,t)=- \ddot{h}_{ij} (0,t)/2$.
We then finally obtain the action for the geodesic deviation
\begin{eqnarray}
S_p \simeq \int_{\gamma_{\tau'}} dt \left[ \frac{m}{2} \dot{\xi}^{i^2} +  \frac{m}{4} \ddot{h}_{ij} (0,t) \xi^i \xi^j \right] \,.
\end{eqnarray}
Notice that, when considering gravitation waves with wavelength smaller than the characteristic separation length $\xi$, an approximation \eqref{metric2} cannot be used.
However, we expect that the effect from such gravitational waves will be suppressed because of the equivalence principle. Hence, we consider the action of the form
\begin{eqnarray}
S_p  \simeq \int_{\gamma_{\tau'}} dt \left[ \frac{m}{2} \dot{\xi}^{i^2} 
+ \frac{m}{2}\frac{\kappa}{\sqrt{V}}\sum_A\sum_{{\bf k}\leq\Omega_{\rm m}} 
\,\ddot{h}^A ({\bf{k}},t )\,e_{ij}^A({\bf k}) \xi^i \xi^j \right] \, ,
\label{action-mirrors}
\end{eqnarray}
where the metric $h_{ij}( 0 , t)$ is replaced  by  the Fourier modes in Eq.~(\ref{fourier}) 
and $\sum_{{\bf k}\leq\Omega_{\rm m}}$  represents the mode sum with the UV cutoff $\Omega_{\rm m}\sim \xi^{-1}$. 
We see a qubic derivative interaction appeared in the above action.

\subsection{Particles in an environment of gravitons}
\label{section2.3}

From Eqs.~(\ref{action-gravitons}) and (\ref{action-mirrors}), the total action $S=S_g+S_p$ we consider is given by
\begin{eqnarray}
S&\simeq&  \int dt \sum_{{\bf k}, A}  \left[ \frac{1}{2} \dot{h}^{A} ({\bf{k}},t)  \dot{h}^{*A}({\bf{k}},t)
 -   \frac{1}{2} k^2 h^A({\bf{k}},t) h^{*A}({\bf{k}},t)  \right]   \nonumber\\
&& \qquad  + \int dt \left[ \frac{m}{2} \dot{\xi}^{i^2}
+  \frac{m}{2}\frac{\kappa}{\sqrt{V}}\sum_A\sum_{{\bf k}\leq\Omega_{\rm m}} 
\,\ddot{h}^A ({\bf{k}},t ) \,e_{ij}^A({\bf k}) \xi^i \xi^j \right] \ .
\label{action-total}
\end{eqnarray}
Note that the geodesic deviation of particles in the graviton background is studied in~\cite{DeLorenci:2014vwa, Parikh:2020nrd, Quinones:2017wka}.
Now we canonically quantize this system. We can expand the interaction picture field $\hat h^A_{\rm I}({\bf k},t)$, whose time  evolution is governed by the quadratic action, in terms of the creation and annihilation operators as 
\begin{align}
\hat h^A_{\rm I}({\bf k},t)
      =\hat a_A({\bf k})u_k(t)+\hat a^\dag_A(-{\bf k})u^*_k(t) \   \,,
\label{hq}
\end{align} 
where the creation and annihilation operators satisfy the standard commutation relations\footnote{$\delta_{{\bf k},{\bf k'}}$ is replaced by $\delta^{(3)}({\bf k}-{\bf k'})$ when taking the infinite volume limit $L_x,L_y,L_z\to\infty$. }
\begin{align}
\left[\,{\hat a}_A({\bf k})\,,\,\hat{a}^\dag_{A'}({\bf k'})\,\right]
&=\delta_{{\bf k},{\bf{k}'}}\delta_{AA'}\,,\qquad
\left[\,{\hat a}_A({\bf k})\,,\,\hat{a}_{A'}({\bf k'})\,\right]=\left[\,{\hat a}^\dag_A({\bf k})\,,\,\hat{a}^\dag_{A'}({\bf k'})\,\right]=0\,                 \label{eq:quantization1}
\end{align}
and $u_k(t)$ denotes a mode function properly normalized as
\begin{align}
\dot u_k(t)u_k^*(t)-u_k(t)\dot u_k^*(t)=-i\,.
\end{align} 
The Minkowski vacuum $\ket{0}$ is defined by $\hat a_A({\bf k}) \ket{0} =0$, with choosing the mode function as 
\begin{align}
u_k(t)=\frac{1}{\sqrt{2k}}e^{-ikt}\equiv u_k^{\rm M}(t)   \ .
\label{minkowski}
\end{align} 

An expectation value of $\hat h^A_{\rm I}({\bf k},t)$ can be non-zero for generic quantum states, such as coherent states as we will see in Section~{\ref{section3.1.1}}. In this case, it may be convenient to divide the gravitational perturbation $\hat h^A_{\rm I}$ around the Minkowski background into a ``classical'' piece 
$h_{\rm cl}({\bf k},t)\equiv\left\langle\hat h^A_{\rm I}({\bf k},t)\right\rangle$ 
and a ``quantum'' piece as
\begin{align}
\delta\hat h^A_{\rm I}({\bf k},t)  = \hat h^A_{\rm I}({\bf k},t) - h_{\rm cl}({\bf k},t)\,.
\end{align}
Here, $\left\langle\hat X\right\rangle$ denotes an expectation value of an operator $\hat X$ for a given quantum state. A precise value of the expectation value depends on the given quantum state. We refer to $\delta\hat h$ as the gravitational quantum fluctuations, {\it i.e.,} gravitons, in the presence of the classical gravitational perturbations $h_{\rm cl}$.
Similarly, we promote the position $\xi^i(t)$ to the operator $\hat{\xi}^i(t)$ below.

\subsection{Langevin type equation of geodesic deviation}
\label{section2.4}

The variation of the action Eq.~(\ref{action-total}) with respect to $h^{*A}$ and $\xi^i$ gives the following equations of motion for the operators in the Heisenberg picture:
\begin{eqnarray}
\ddot{\hat{h}}^A({\bf k} ,t) + k^2 \hat{h}^A({\bf k} ,t)  &=& \frac{\kappa m}{2\sqrt{V}}  e^{*A}_{ij} ({\bf k})  \frac{d^2}{dt^2} \left\{\hat{\xi}^i (t) \hat{\xi}^j (t) \right\} 
\label{gw_eq} \,,\\
\ddot{\hat{\xi}}^i (t) &=&  \frac{\kappa}{\sqrt{V}}  \sum_A\sum_{{\bf k}\leq\Omega_{\rm m}}  e^{A}_{ij} ({\bf k})  \ddot{\hat{h}}^A ({\bf k} ,t) \hat{\xi}^j  (t)   
 \label{dev_eq}  \,.
\end{eqnarray}
Eq.(\ref{gw_eq}) is solved by standard Green's function techniques. Specifically we consider the setup where the interaction between $\hat h$ and $\xi^i$ is turned on at $t=0$. Under this setup, the formal solution of Eq.~\eqref{gw_eq} is  given by
\begin{align}
\hat{h}^A({\bf k} ,t) &=& h_{\rm cl}^A({\bf k},t)+ \delta\hat{h}^A_{\rm I}({\bf k},t )
+    \frac{\kappa m}{2\sqrt{V}}  e^{*A}_{ij} ({\bf k})  \int_0^t dt'   \frac{\sin k(t-t')}{k}  \frac{d^2}{dt^{\prime 2}} \left\{\hat{\xi}^i (t') \hat{\xi}^j (t') \right\}     \ .
 \label{green}
\end{align}
The last nonhomogeneous solution describes the gravitational waves emitted from the particles. 
Substituting the formal solution Eq.~(\ref{green}) into Eq.~(\ref{dev_eq}), we have
\begin{eqnarray}
\ddot{\xi^i}(t)&=& \frac{1}{2} \ddot{h}^{\rm cl}_{ij}  (0 ,t) \hat{\xi}^j (t)
   -\frac{\kappa}{\sqrt{V}}  \xi^j (t) \sum_A\sum_{{\bf k}\leq\Omega_{\rm m}}  k^2  e^{A}_{ij}\, \delta\hat h^A_{\rm I}({\bf k},t) \nonumber\\
&&    -  \frac{\kappa^2 m}{4V}  \xi^j (t) \sum_{{\bf k}\leq\Omega_{\rm m}} \left[ P_{ik} P_{j\ell} +  P_{i\ell} P_{jk} - P_{ij}  P_{k\ell}      \right]   \int_0^t dt'   k\sin k(t-t')  \frac{d^2}{dt^{\prime 2}} \left\{\hat{\xi}^k (t') \hat{\xi}^\ell (t') \right\}   \nonumber \\
&&    +   \frac{\kappa^2 m}{4V}  \sum_{{\bf k}\leq\Omega_{\rm m}} \left[ P_{ik} P_{j\ell} +  P_{i\ell} P_{jk} - P_{ij}  P_{k\ell}      \right] \hat{\xi}^j (t) \frac{d^2}{dt^{ 2}} \left\{\hat{\xi}^k (t) \hat{\xi}^\ell (t) \right\}  \ ,
\label{eom:xi}
\end{eqnarray}
where we have defined 
\begin{eqnarray}
h^{\rm cl}_{ ij} (x^i,t) \equiv \frac{2\kappa}{\sqrt{V}}\sum_{{\bf k},A}e_{ij}^A({\bf k}) h_{\rm cl}^A({\bf k},t)\,e^{i\bf{k}\cdot\bf{x}}
\ ,
\label{hc}
\end{eqnarray}
and introduced the projection tensor $P_{ij}  = \delta_{ij} -\bar{k}_i \bar{k}_j$ orthogonal to the unit wave number $\bar{k}_i =k_i /k$ and used
\begin{eqnarray}
\sum_A e^{*A}_{ij}({\bf k})e^A_{k\ell}({\bf k}) = \frac{1}{2}\left[ P_{ik} P_{j\ell} +  P_{i\ell} P_{jk} - P_{ij}  P_{k\ell}      \right]  \ . 
\label{eq:polsum}
\end{eqnarray}
Here, the UV-regulated mode sum $\sum_{{\bf k}\leq\Omega_{\rm m}}$ in the second and the third line of \eqref{eom:xi} can be performed by taking the continuum limit of the {\bf k}-mode by removing the width introduced in Eq.~(\ref{fourier}): $1/V\sum_{{\bf k}\leq\Omega_{\rm m}} \rightarrow 1/(2\pi)^3\int ^{\Omega_{\rm m}}d^3 k$. The momentum integral is computed in Appendix~\ref{appA}, and the result is
\begin{eqnarray}
&& \ddot{\hat{\xi}}^i (t) -\frac{1}{2} \ddot{h}^{\rm cl}_{ ij} (0,t) \hat{ \xi}^j(t)+   \frac{\kappa^2 m }{40\pi}    \left[ \delta_{ik} \delta_{j\ell} +  \delta_{i\ell} \delta_{jk} - \frac{2}{3}\delta_{ij}  \delta_{k\ell}      \right] 
           \hat{\xi}^j (t) \frac{d^5}{dt^{ 5}} \left\{\hat{\xi}^k (t) \hat{\xi}^\ell (t) \right\}         \nonumber  \\
&& \quad =  -\delta\hat{N}_{ij}(t)\hat{\xi}^j (t)    +   \frac{\kappa^2 m}{20\pi^2}  \Omega_{\rm m} \left[ \delta_{ik} \delta_{j\ell} +  \delta_{i\ell} \delta_{jk} - \frac{2}{3}\delta_{ij}  \delta_{k\ell}      \right] 
\hat{\xi}^j (t) \frac{d^4}{dt^{ 4}} \left\{\hat{\xi}^k (t) \hat{\xi}^\ell (t) \right\}  
\label{langevin}
\end{eqnarray}
where we have defined
\begin{eqnarray}
\delta\hat{N}_{ij}(t)\equiv\frac{\kappa}{\sqrt{V}} \sum_A \sum_{{\bf k}\leq\Omega_{\rm m}} k^2  e^{A}_{ij}({\bf k} ) \delta\hat h^A_{\rm I}({\bf k},t)\ . \label{eq:noise}
\end{eqnarray}
The third term on the left hand side represents the force of radiation reaction. On the right hand side, 
the first term is the random force induced by gravitons, which is nothing but quantum noise. 
The last term is not relevant in the discussions of the noise and the decoherence. The schematic expression of Eq.~(\ref{langevin}) without the last term is obtained in~\cite{Parikh:2020nrd}.
The noise of gravitons $\delta \hat N_{ij}$ always exists if the gravitational waves are quantized. If the noise is detected, it would be an evidence that gravity is quantized.

\section{Quantum noise induced by gravitons}
\label{section3}
\subsection{Quantum noise correlations}
\label{section3.1}
In this section, we compute the amplitude of the noise for various quantum states. 

The anticommutator correlation function of $\delta N_{ij}(t)$ can be computed by using Eqs.~\eqref{eq:polsum} and \eqref{eq:quantization1} in the infinite volume limit $L_x,L_y,L_z\to\infty$ as
\begin{align}
\left<\left\{\delta \hat N_{ij}(t),\,\delta \hat N_{k\ell}(t')\right\}\right> =\frac{\kappa^2}{10\pi^2} \left( \delta_{ik} \delta_{j\ell} +  \delta_{i\ell} \delta_{jk} 
                 - \frac{2}{3}  \delta_{ij}  \delta_{k\ell}      \right)\int^{\Omega_{\rm m}}_0\mathrm{d}k\,k^6P_{\delta h}(k,t,t')\,,\label{eq:noiseamp1}
\end{align}
where we defined the anticommutator symbol $\{\cdot\,,\cdot\}$ as $\{\hat X,\hat Y\}\equiv(\hat X\hat Y+\hat Y\hat X)/2$, and  $P_{\delta h}$ is given by 
\begin{align}
\left<\left\{\delta h^A_{\rm I}({\bf k},t),\,\delta h^{A'}_{\rm I}({\bf k'},t')\right\}\right>
=\delta_{AA'}\delta_{{\bf k}+{\bf k'},{\bf 0}}P_{\delta h}(k,t,t')\,.\label{eq:consv1}
\end{align}
Below we compute the noise correlation functions when the graviton is in a squeezed-coherent state and discuss the Minkowski vacuum as a special case. The  coherent state or squeezed state can be realized when the squeezing parameter or the coherent parameter goes to zero, respectively.

\subsubsection{Squeezed coherent states}
\label{section3.1.1}
The definition of the squeezed coherent state $\ket{\zeta,B}$ is
\begin{align}
&\ket{\zeta,B}\equiv \hat S(\zeta)\hat D(B)\ket{0}\,,
\end{align}
where $\hat S(\zeta)$ and $\hat D(B)$ are the squeezing and the displacement operators, respectively. 
They are expressed by
\begin{eqnarray}
&&\hat S(\zeta)\equiv \exp\left[\frac{1}{V}\sum_{{\bf k},A}
\left(\zeta^*_k\,\hat a_A({\bf k})\hat a_A(-{\bf k})
+ \zeta_k \,{{\hat a}_A^\dag}  ({\bf k}) {\hat a}_A^\dag (-{\bf k})  \right) \right]  \,,     \\
&&\hat D(B)\equiv \exp\left[\frac{1}{V}\sum_{{\bf k},A}\left(B_k\,\hat a^\dag_A({\bf k})-B_k^*\,\hat a_A({\bf k})\right)\right]\,,
\end{eqnarray}
where $\zeta_k\equiv r_k\exp[i\varphi_k]$ and $r_k$ is the squeezing parameter. $B_k$ is the coherent parameter. We assume that the parameter $\zeta_k$ and $B_k$ only depend on $k$ and are independent of the direction of $\bf k$. These operators are unitary, and satisfy the following relations:
\begin{eqnarray}
&&\hspace{-1.5cm}
\hat D^\dag(B)\hat S^\dag(\zeta)\,\hat a_A({\bf k})\,\hat S(\zeta)\hat D(B)
=\left(\hat a_A({\bf k})+B_k\right)\cosh r_k
-\left(\hat a_A^\dag(-{\bf k})+B^*_k\right)e^{i\varphi_k}\sinh r_k\,,\\
&&\hspace{-1.5cm}
\hat D^\dag(B)\hat S^\dag(\zeta)\,\hat a_A^\dag({-\bf k})\,\hat S(\zeta)\hat D(B)
=\left(\hat a_A^\dag({-\bf k})+B_k^*\right)\cosh r_k
-\left(\hat a_A({\bf k})+B_k\right)e^{-i\varphi_k}\sinh r_k\,.
\label{eq:transf1}
\end{eqnarray}
The vacuum expectation value of the above operators become
\begin{eqnarray}
&&\hspace{-1.5cm}
\langle 0|\hat D^\dag(B)\hat S^\dag(\zeta)\,\hat a_A({\bf k})\,\hat S(\zeta)\hat D(B)|0\rangle
=B_k\cosh r_k-B^*_k e^{i\varphi_k}\sinh r_k\,,\\
&&\hspace{-1.5cm}
\langle 0|\hat D^\dag(B)\hat S^\dag(\zeta)\,\hat a_A^\dag({-\bf k})\,\hat S(\zeta)\hat D(B)|0\rangle
=B_k^*\cosh r_k
-B_k e^{-i\varphi_k}\sinh r_k\,,
\end{eqnarray}
On the other hand, the transformation of the operator $\delta\hat{h}^A_{\rm I}({\bf k},t)$ is given by definition as
\begin{eqnarray}
&&
\hat D^\dag(B)\hat S^\dag(\zeta)\,\delta\hat{h}^A_{\rm I}({\bf k},t)\,\hat S(\zeta)\hat D(B)
=\hat D^\dag(B)\hat S^\dag(\zeta)\,\hat{h}^A_{\rm I}({\bf k},t)\,\hat S(\zeta)\hat D(B)\nonumber\\
&&\hspace{6.5cm}
-\langle 0|\hat D^\dag(B)\hat S^\dag(\zeta)\,\hat{h}^A_{\rm I}({\bf k},t)\,\hat S(\zeta)\hat D(B)
|0\rangle\,.
\end{eqnarray}
Because $\hat{h}^A_{\rm I}({\bf k},t)$ consists of the $\hat a_A({\bf k})$ and $\hat a_A^\dag({\bf k})$, we see that the right hand side of the above relation is independent of the coherent parameter $B_k$ and we have
\begin{eqnarray}
&&\hspace{-1.5cm}
\hat D^\dag(B)\hat S^\dag(\zeta)\,\delta\hat{h}^A_{\rm I}({\bf k},t)\,\hat S(\zeta)\hat D(B)
=\left(\hat a_A({\bf k})\cosh r_k
-\hat a_A^\dag(-{\bf k})e^{i\varphi_k}\sinh r_k\right)u_k(t)\nonumber\\
&&\hspace{5cm}
+\left(\hat a_A^\dag({-\bf k})\cosh r_k
-\hat a_A({\bf k})e^{-i\varphi_k}\sinh r_k\right)u^*_k(t)\nonumber\\
&&\hspace{4.2cm}
=u^{\rm sq}_k(t)\hat a_A({\bf k})+{u^{\rm sq}_k}^*(t)\hat a_A^\dag(-{\bf k})\nonumber\\
&&\hspace{4.2cm}
=\hat S^\dag(\zeta)\,   \hat{h}^A_{\rm I}({\bf k},t)\,\hat S(\zeta)\,,
\end{eqnarray}
where the mode function in the squeezed state is given in terms of that in the Minkowski space in Eq.~(\ref{minkowski}) such as
\begin{eqnarray}
u^{\rm sq}_k(t)\equiv u^{\rm M}_k(t)\cosh r_k-e^{-i\varphi_k}{u_k^{\rm M}}^*(t)\sinh r_k\,.
\end{eqnarray}
Hence, 
the anticommutator correlation function of $\delta \hat N_{ij}(t)$ in the squeezed coherent state  in Eq.~(\ref{eq:noiseamp1}) becomes independent of the coherent state such as
\begin{eqnarray}
&& \Big\langle \zeta,B\Big|\left\{\delta \hat N_{ij}(t),\,\delta \hat N_{k\ell}(t')\right\}\Big|\zeta, B\Big\rangle 
\nonumber \\
&& \qquad =\frac{\kappa^2}{10\pi^2} \left( \delta_{ik} \delta_{j\ell} +  \delta_{i\ell} \delta_{jk} 
                 - \frac{2}{3}  \delta_{ij}  \delta_{k\ell}      \right)\int^{\Omega_m}_0\mathrm{d}k\,k^6\mathrm{Re}\left[u^{\rm sq}_k(t){u^{\rm sq}_k}^*(t')\right]\,, 
\label{eq:noiseamp3}
\end{eqnarray}
where
\begin{eqnarray}
P_{\delta h}(k,t,t')&=&
\mathrm{Re}\left[u^{\rm sq}_k(t){u^{\rm sq}_k}^*(t')\right]\nonumber\\
&=&\frac{1}{2k}\Bigl[\cos\{k(t-t')\}\cosh 2r_k-\cos\{k(t-t')-\varphi_k\}\sinh 2r_k\Bigr] \ .\label{eq:squezamp}
\end{eqnarray}
In general, the squeezing parameter $r_k$ and the phase $\varphi_k$ depend on $k$. 
However, for simplicity, we regard these variables as constants. Then,  plugging this into Eq.~(\ref{eq:noiseamp3}), we obtain
\begin{eqnarray}
&& \Big\langle\zeta,B\Big|\left\{\delta \hat N_{ij}(t),\,\delta \hat N_{k\ell}(t')\right\}\Big|\zeta,B\Big\rangle 
\nonumber\\
&&\qquad =\frac{\kappa^2\Omega_{m}^6}{20\pi^2} \left( \delta_{ik} \delta_{j\ell} +  \delta_{i\ell} \delta_{jk} 
 - \frac{2}{3}  \delta_{ij}  \delta_{k\ell}\right) F(\Omega_m(t-t'),r,\varphi)\,,
 \label{deltaN1}
\end{eqnarray}
where
\begin{eqnarray}
&&\hspace{-5mm}
F(x,r,\varphi)=\frac{1}{x^6}\left[\big\{\!\!\left(5x^4-60x^2+120\right)\cos x+x\left(x^4-20x^2+120\right)\sin x-120\big\}\cosh 2r\right.\nonumber\\
&&\left.\hspace{2.5cm}
-\big\{\!\!\left(5x^4-60x^2+120\right)\cos\left(\varphi-x\right)-x\left(x^4-20x^2+120\right)\sin\left(\varphi-x\right)\right.\nonumber\\
&&\left.\hspace{2.5cm}-120\cos\varphi\big\}\sinh 2r\right]\,.
\label{F0}
\end{eqnarray}
Note that $F(x,r,\varphi)$ converges to zero for large $x$ and the function  $F(x,r,\varphi)$  for small $x$ can be expanded as
\begin{eqnarray}
F(x,r,\varphi)&=&\frac{1}{6}\left(\cosh 2r-\cos\varphi\sinh 2r\right)-\frac{10}{7}\sin\varphi\sinh 2r\, x
\nonumber\\
&&  \qquad -\frac{25}{4}\left(\cosh 2r-\cos\varphi\sinh 2r\right) x^2+\mathcal{O}\left(x^3\right)\,.
\label{F1}
\end{eqnarray}
We find that quantum noise correlations increases as $\Omega_{\rm m}$ increases. 

\subsubsection{The Minkowski vacuum state}
\label{section3.1.2}
For comparison, let us see the correlation functions of the quantum noise in the Minkowski vacuum state which is obtained by taking  $r_k\rightarrow 0$ and then we have
\begin{align}
P_{\delta h}(k,t,t')=\mathrm{Re}\left[u^{\rm M}_k(t){u^{\rm M}_k}^*(t')\right]=\frac{1}{2k}\cos\{k(t-t')\}\,.
\end{align}
Substituting this into Eq.~\eqref{eq:noiseamp1}, we get 
\begin{eqnarray}
\label{eq:noiseamp2}
\Big\langle 0\Big|\left\{\delta \hat N_{ij}(t),\,\delta \hat N_{k\ell}(t')\right\}\Big|0\Big\rangle =\frac{\kappa^2\Omega_{m}^6}{20\pi^2} \left( \delta_{ik} \delta_{j\ell} +  \delta_{i\ell} \delta_{jk} 
                 - \frac{2}{3}  \delta_{ij}  \delta_{k\ell}      \right) F(\Omega_m(t-t'))\,, 
\end{eqnarray}
where
\begin{eqnarray}
F(x)\equiv\frac{1}{x^6}\left[\left(5x^4-60x^2+120\right)\cos x+x\left(x^4-20x^2+120\right)\sin x-120\right]\,.
\end{eqnarray}
Note that, for small $x$, the function  $F(x)$ can be expanded as
\begin{eqnarray}
F(x)=\frac{1}{6}-\frac{25}{4} x^2+\mathcal{O}\left(x^4\right)\,.
\label{F1}
\end{eqnarray}
Comparing this with the result of squeezed coherent state, we see the quantum noise correlations are enhanced exponentially by the squeezing parameter.

\subsection{Detectability of the quantum noise}
\label{section3.2}
In this subsection, we roughly estimate the effective strain $h_{\rm eff}$ corresponding to the quantum noise $\delta\hat{N}_{ij}$ and discuss the detectability of the quantum noise. For a given quantum state, the amplitude of the quantum noise in frequency domain can be characterized as
\begin{align}
\delta N(f)&\equiv\left(\int^\infty_{-\infty}\mathrm{d}t\,\left<\{\delta N^{ij}(t),\delta N_{ij}(0)\}\right>e^{2\pi ift}\right)^{\frac{1}{2}}\,.
\end{align}
From Eq.~\eqref{langevin}, it is found that the response of $\xi^i$ to presence of the classical gravitational wave and the quautum noise is proportional to $\ddot h_{\rm cl}(t,0)$ and $\delta N(t)$, respectively. Here we omitted spatial indices. This suggests that we can discuss the detectablity of the noise by using the effective strain $h_{\rm eff}(f)\equiv (2\pi f)^{-2}\delta N(f)$ in frequency domain.

Let us start with Minkowski vacuum state. In this case, the amplitude of the quantum noise in frequency domain can be computed as
\begin{align}
\delta N(f)=\pi^2\frac{(2f)^{\frac{5}{2}}}{M_{\rm pl}}\,,
\label{deltaN}
\end{align}
where the reduced Planck mass is $M_{\rm pl}\sim 10^{18}$ GeV.  Corresponding effective strain is then
\begin{eqnarray}
h_{\rm eff}(f)=\frac{(2f)^{\frac{1}{2}}}{M_{\rm pl}}\approx 2\times 10^{-42} \left( \frac{f}{1{\rm Hz}} \right)^{\frac{1}{2}}\quad{\rm Hz}^{-\frac{1}{2}}  \ .
\end{eqnarray}
For instance, the characteristic frequency of LIGO is around $100$ Hz. Then the amplitude of quantum noise becomes $h_{\rm eff}(f)|_{f\sim100\,{\rm Hz}}\sim 10^{-41}\,{\rm Hz}^{-1/2}$. The strain sensitivity of LIGO is about $10^{-23}\,{\rm Hz}^{-1/2}$ for $f\sim 100$ Hz, so the amplitude of quantum noise is too small to be detected.

However, if gravitational waves are in the squeezed state (or in the squeezed coherent state) when arriving at the detectors, the amplitude of the quantum noise is enhanced by the exponential factor of squeezing parameter as seen in Eq.~\eqref{eq:squezamp}. That is,
\begin{eqnarray}
   h_{\rm eff}(f)   \approx 2\times 
   10^{-42} \left( \frac{f}{1{\rm Hz}} \right)^{\frac{1}{2}}
   e^{r_k}|_{k=2\pi f}\quad{\rm Hz}^{-\frac{1}{2}}
   \,.\label{eq:sqstrain}
\end{eqnarray}
For instance, if 
the squeezing parameter is large as much as $e^{r_k} \sim 10^{22}$, the amplitude of the quantum noise at the characteristic frequency of LIGO becomes $h_{\rm eff}(f)   \sim 
   10^{-20}\,{\rm Hz}^{-\frac{1}{2}}$, which is detectably large. This point is emphasized in~\cite{Parikh:2020nrd}.


One possible and well-known mechanism to produce gravitons with large quantum fluctuations is inflation. Gravitons produced during inflation experience large squeezing which leads to the detectably large noise amplitudes.\footnote{The gravitons may undergo the quantum decoherence during inflation or during the propagation. However, when discussing the amplitude of the noise, we do not need to take care of the decoherence.} In the case of primordial gravitational waves, the relation between the squeezing parameter and the current frequency $f$ is given by
\begin{eqnarray}
    e^{r_k}|_{k=2\pi f}  \approx  \left( \frac{f_{\rm c}}{f} \right)^2
   \ ,
\end{eqnarray}
where $f_{\rm c}$ is the cutoff frequency. In the case of GUT inflation, we have $f_{\rm c} \sim 10^8 $ Hz. 
In this case, the effective strain at $f\sim 0.1$ Hz, which is the characteristic frequency of DECIGO\footnote{DECIGO stands for DECi-hertz Interferometer Gravitational wave Observatory. It is a gravitational wave antenna in space operating in the 0.1 - 10 Hz frequency band. It consists of three drag-free spacecraft, 1,000 km apart from each other. }~\cite{Seto:2001qf,Kawamura:2011zz}, reads $h_{\rm eff}(f)   \sim 
10^{-24}\,{\rm Hz}^{-\frac{1}{2}}$
because we have $e^{r_k}\sim 10^{18}$.
This is the reason why the stochastic gravitational waves from inflation could be detectably large.
We here stress that, if we could detect the noise $\delta\hat N_{ij}$, it would imply discovery of gravitons.

\section{Decoherence induced by gravitons}
\label{section4}

In the previous section, we discussed the effect of noise induced by gravitons on gravitational wave detectors. In this section, we explore the effect of gravitons on the process of  decoherence in laboratory. 

\begin{figure}[t]
\vspace{-1cm}
\includegraphics[height=7cm]{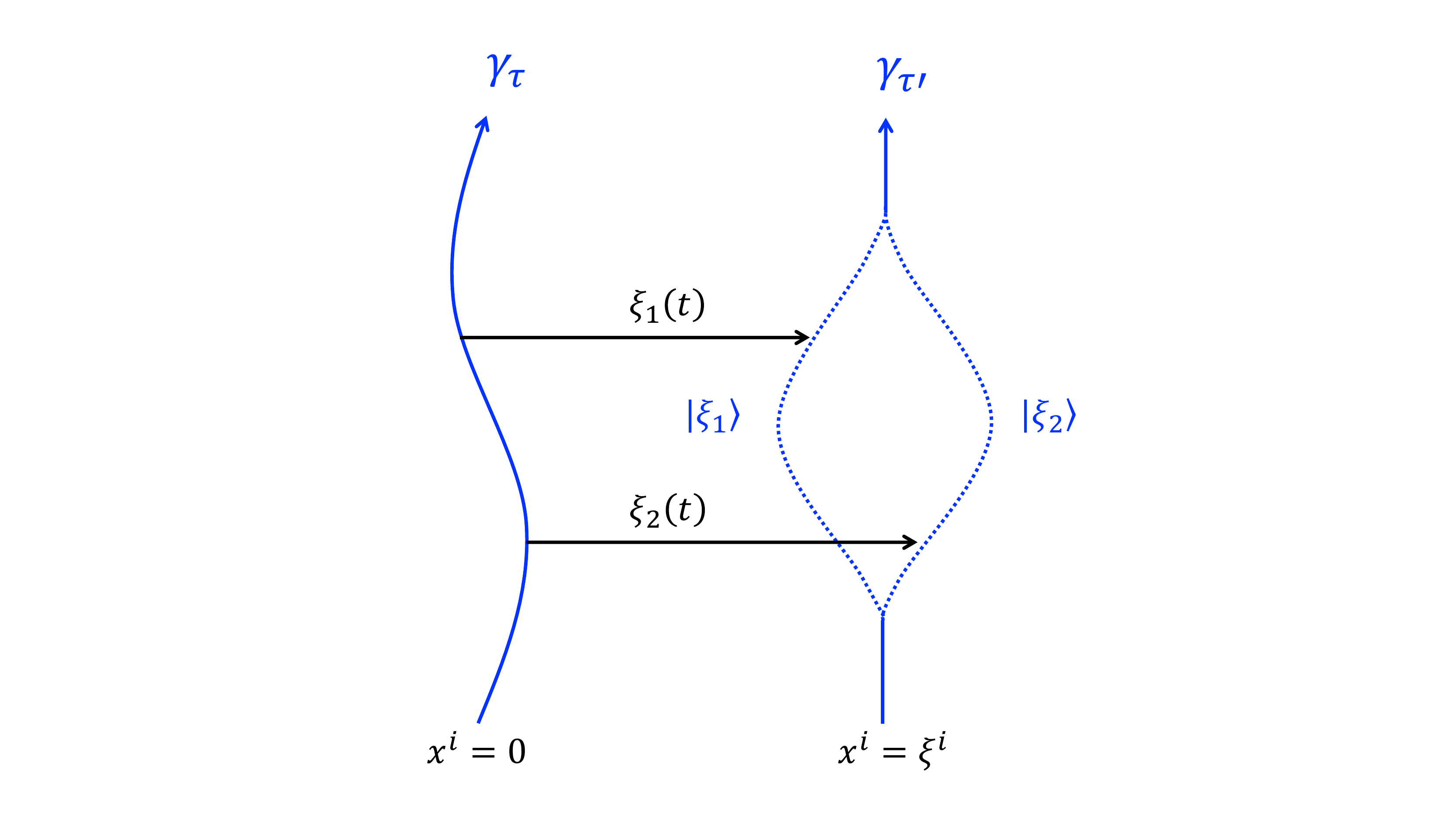}\centering
\vspace{-0.5cm}
\caption{Timelike curves of two massive particles $\gamma_\tau$ and $\gamma_{\tau'}$, and the $\gamma_{\tau'}$ is in a superposition state of two spatially-separated locations $\xi_1(t)$ and $\xi_2(t)$.  The superposition state survives $\xi_1(t)\neq\xi_2(t)$ for the duration $0< t< t_{\rm f}$.} 
\label{fig2}
\end{figure}

We consider a system of two massive particles, one of which is in a superposition state of two spatially-separated locations as shown in Figure~\ref{fig2}, and investigate the loss of coherence of the  superposition state caused by gravitons. If the decoherence process due to gravitons is detectable in laboratory, it would be strong evidence that gravity is quantized.
\footnote{Strictly speaking, we may also need to take into account the intrinsic gravitational decoherence proposed by~\cite{Diosi:1986nu,Penrose:1996cv}.}

\subsection{Setup and the decoherence functional}\label{subsec:gravity}
We consider timelike curves $\gamma_{\tau}$ and $\gamma_{\tau'}$ of two massive particles with a mass $m$ and $\gamma_{\tau'}$ is spatially superposed across the distance $\xi_2-\xi_1$ in the Minkowski space (See  Figure~\ref{fig2}). The superposition state survives for $0<t<t_{\rm f}$, and the system is assumed to be in a certain initial quantum state $\ket{\Psi(t_0)}$ at the initial time $t_0<0$. The normalization condition is $\langle\Psi(t_0)|\Psi(t_0)\rangle=1$. 

In the Schr\"{o}dinger picture, the unitary time evolution of the superposition state is expressed as
\begin{eqnarray}
\ket{\Psi(t_0)}\rightarrow|\Psi(t)\rangle=|\xi_1(t)\rangle+|\xi_2(t)\rangle\,,
\end{eqnarray}
where $|\xi_1(t)\rangle$ and $|\xi_2(t)\rangle$ are approximately eigenstates of the operator $\hat{\xi}^i$ satisfying $\hat{\xi}^i|\xi_1(t)\rangle=\xi_1^i(t)|\xi_1(t)\rangle$ and  $\hat{\xi}^i|\xi_2(t)\rangle=\xi_2^i(t)|\xi_2(t)\rangle$. 
The superposition state lasts for $0<t<t_{\rm f}$ and
$\xi^i_1(t)=\xi^i_2(t)$ holds for $t\notin(0,t_{\rm f})$. 

The decoherence arises through the interaction of the system of massive particles with the environmental gravitons in this setup. The leading interaction is given in the last term on the right hand side of \eqref{action-total}:
\begin{align}
S\supset  \frac{m}{2}\int dt \frac{\kappa}{\sqrt{V}}\sum_A\sum_{{\bf k}\leq\Omega_{\rm m}} 
\,\ddot{h}^A ({\bf{k}},t ) \,e_{ij}^A({\bf k}) \xi^i \xi^j\,.
\end{align}
To discuss the rate of decoherence, we use the influence functional method~\cite{Feynman:1963fq} and compute the modulus of the influence functional
\begin{align}
\exp[-\Gamma(t_{\rm f})]\equiv\left|\frac{\left<\xi_2(t_{\rm f})\right|\left.\xi_1(t_{\rm f})\right>}{\left<\xi_2(t_0)\right|\left.\xi_1(t_0)\right>}\right|\,.\label{eq:decofn1}
\end{align}
The loss of quantum coherence between $\ket{\xi_1(t)}$ and $\ket{\xi_2(t)}$ occurs when $\exp[-\Gamma(t_{\rm f})]\ll1$.
We can compute the decoherence functional by integrating out the gravitons as\footnote{For the derivation of \eqref{rate}, see Appendix~\ref{app:infl}.}
\begin{align}
\Gamma (\tf) \approx\frac{m^2}{8}\int^{\tf}_0\mathrm{d}t\,\Delta(\xi^i\xi^j)(t)\int^{\tf}_0\mathrm{d}t'\,\Delta(\xi^k\xi^\ell)(t')\Big\langle\left\{\delta\hat N_{ij}(t),\,\delta\hat N_{k\ell}(t')\right\}\Big\rangle\,.
\label{rate}
\end{align}
Here $\Delta(\xi^i\xi^j)(t)=\xi^i_1(t)\xi^j_1(t)-\xi^i_2(t)\xi^j_2(t)$ denotes a difference of $\xi^i(t)\xi^j(t)$ in the superposition. The value of $\Delta(\xi^i\xi^j)(t)$ is determined by an experimental setup. We consider some simple configurations of the superposition state in order to evaluate the rate of decoherence explicitly in the next subsection.

\subsection{Decoherence rate for simple configurations of the superposition state}
\label{subsec:rate}
For simplicity, we assume that $\xi^i_a(t)=\xi_a(t)\delta^{i1}$ for $a=1,2$ and that $\xi\equiv(\xi_1(t)+\xi_2(t))/2$ is independent of time. We then consider the configuration of the superposition state with the separation $\Delta\xi(t)\equiv\xi_2(t)-\xi_1(t)$ as\footnote{This parameterization is essentially the same as the one in \cite{Breuer and Petruccione}.} 
\begin{equation}
 \Delta \xi(t) =
\begin{cases}
 2vt & {\rm for} \quad 0<t\leq \tf/2 \,,\\
 2v(\tf-t) & {\rm for} \quad \tf/2<t<\tf \,,\label{eq:model}
\end{cases} 
\end{equation}
where the norm of velocity $v$ is constant ($0<v<1$). The velocity of the particle changes only at the moment $t=\tf/2$ in this configuration. 
\begin{figure}[tbp]
 \centering
  \includegraphics[width=.5\textwidth, trim=170 100 150 150,clip]{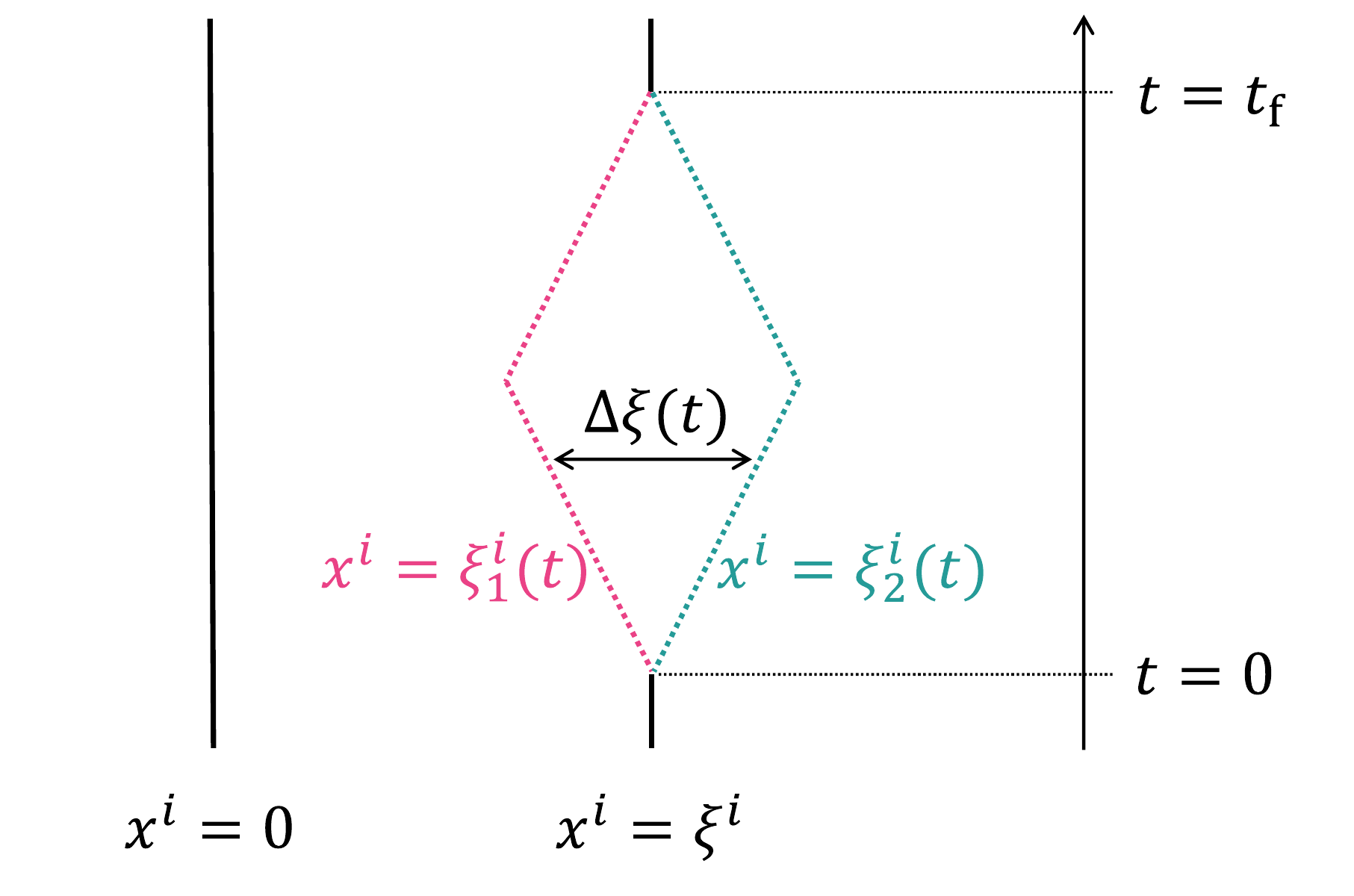}
 \caption{The schematic diagram of the configuration of the superposition state given in Eq.~\eqref{eq:model}. 
 The trajectory at $x^i=\xi^i$ is spatially superposed across the distance $\Delta\xi(t)$ for the duration $0<t<\tf$. The superposition state consists of the dotted red and green lines. The averaged trajectory of the two dotted lines $\xi$ is time independent.}
 \label{fig:model} 
\end{figure}
The schematic diagram of this configuration is depicted in Figure~\ref{fig:model}. The maximum  separation at $t=\tf/2$ is $\Delta\xi(\tf/2)=v\tf\equiv\dl$.

\begin{figure}[t]
 \centering
  \includegraphics[width=.35\textwidth, trim=70 390 30 60,clip]{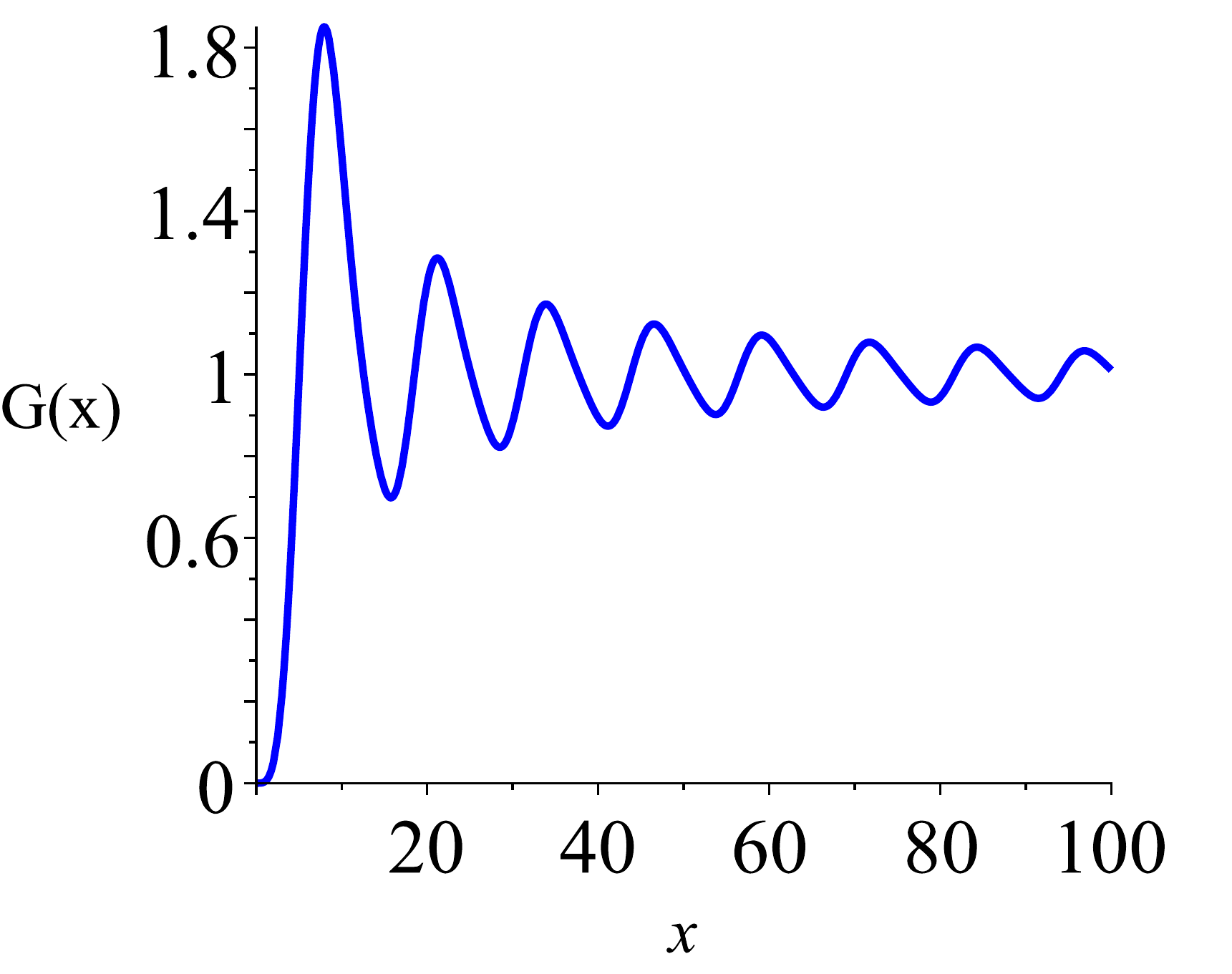}
 \caption{A plot of $G(x)$. $G(x)$ grows polynomially for $x\ll1$ and then oscillatory behavior converges to order unity for $x\gtrsim10$.}
 \label{fig:plot1} 
\end{figure}

The decoherence rate Eq.~\eqref{rate} in the Minkowski vacuum state is expressed as
\begin{align}
\Gamma(\tf)\approx\frac{m^2\uv^6\xi^2}{30\pi^2\Mpl^2}\int^{\tf}_0\mathrm{d}t\int^{\tf}_0\mathrm{d}t'\Delta\xi(t)\Delta\xi(t')F\left(\uv(t-t')\right)\,.\label{rate2}
\end{align}
Here we used Eq.~\eqref{eq:noiseamp2}. Substituting Eq.~\eqref{eq:model} into Eq.~\eqref{rate2}, we obtain\footnote{In terms of the maximum separation length $\dl$, eq.~\eqref{rate3} can be written as 
\begin{align}
\Gamma(\tf)\approx \frac{2m^2}{5\pi^2\Mpl^2}\left(\uv\xi\right)^2\left(\uv\dl\right)^2\frac{G\left(\uv\tf\right)}{\left(\uv\tf\right)^2}\label{rate4}\,.
\end{align}
Note that $0<\dl<\tf$ because the velocity of the particle cannot exceed the speed of light.}
\begin{align}
\Gamma(\tf)\approx \frac{2m^2v^2}{5\pi^2\Mpl^2}\left(\uv\xi\right)^2G\left(\uv\tf\right)\label{rate3}\,,
\end{align}
where $G(x)$ is given by
\begin{align}
G\left(x\right)\equiv1+\frac{2}{3x}\left[\sin(x)-8\sin \left( \frac{x}{2} \right) \right]+\frac{1}{x^2}\left[\frac{2}{3}\cos(x)-\frac{32}{3}\,\cos \left( \frac{x}{2} \right) +10\right]\,.
\end{align}
The behavior of $G(x)$ is shown in Figure.~\ref{fig:plot1}. The function $G(x)$ grows polynomially as $G(x)\simeq x^4/288+\mathcal{O}(x^6)$ for $x\ll1$ and
shows a damping oscillation around $G(x)=1$ for large $x$.
As we explained in Section~\ref{section2.2}, the cutoff $\uv$ satisfies $\uv\xi\sim\mathcal{O}(1)$. If we take $\uv\xi=1$, then Eq.~\eqref{rate3} reads 
\begin{align}
\Gamma(\tf)\approx \frac{2m^2v^2}{5\pi^2\Mpl^2}G\left(\uv\tf\right)\label{rate5}\,.
\end{align}
Since $G(x)\sim\mathcal{O}(1)$, we find $\Gamma(\tf)\ll1$ as long as the momentum of the massive particle is much smaller than the Planck mass, $mv\ll\Mpl\approx 4\times10^{-6}$ g, and then decoherence does not occur. 
On the other hand, for $mv\gg\Mpl$,  
the decoherence happens. Let us estimate the decoherence time $\tdec$ defined by $\Gamma(\tdec)=1$ in this case.  In order to hold $\Gamma(\tdec)=1$, $G(x)$ has to be less than unity. Thus the decoherence time can be computed by using $G(x)\simeq x^4/288$ $(x\ll 1)$ as
\begin{align}
\uv\tdec\sim 10\times\left(\frac{\Mpl}{mv}\right)^{1/2}\,.
\end{align}
We see the decoherence time is $\uv\tdec\ll 1$ for the case that the momentum of the massive particle is much larger than the Planck mass $mv\gg\Mpl$.

As another simple configuration of the superposition state, we consider a system that the previous configuration Eq.~\eqref{eq:model} repeats $N$ times  as
\begin{align}
\Delta\xi(t)=\sum_{k=0}^{N-1}\Delta\xi(t;k)\,,\label{eq:model2a}
\end{align}
with
\begin{equation}
 \Delta \xi(t;k) \equiv
\begin{cases}
 2vt & {\rm for} \quad k\left(\Ts+\Tc\right)<t\leq (k+1/2)\Ts+k\Tc\,,\\
 2v(\tf-t) & {\rm for} \quad  (k+1/2)\Ts+k\Tc<t<(k+1)\Ts+k\Tc\,,\\
0 & {\rm for}\quad t\leq k\left(\Ts+\Tc\right)\,,\,(k+1)\Ts+k\Tc\leq t\,.\label{eq:model2b}
\end{cases} 
\end{equation}
\begin{figure}[tbp]
 \centering
  \includegraphics[width=.7\textwidth, trim=100 160 210 200,clip]{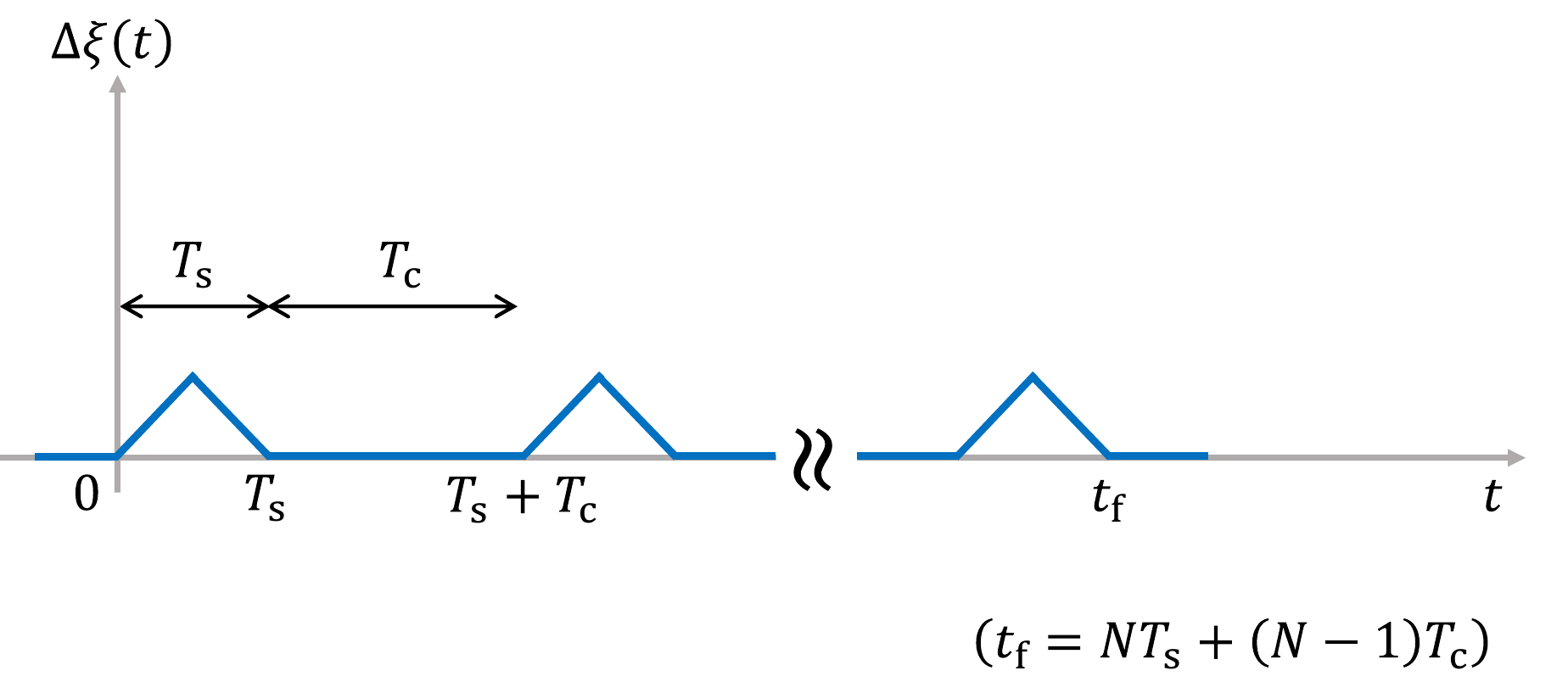}
 \caption{The separation of the configuration of the superposition state as a function of time. The blue solid line indicates the configuration given in Eq.~\eqref{eq:model2a}. The configuration of Eq.~\eqref{eq:model} repeats $N$ times as time evolves. The duration of the superposition state is $\Ts$ and each configuration repeats in particular time interval $\Tc$.}
 \label{fig:model2} 
\end{figure}
The configuration of the superposition state of \eqref{eq:model2a} is shown in Figure~\ref{fig:model2}. 
 The duration of the superposition state is $\Ts$ and each configuration repeats with a time interval $\Tc$.
We assume $\Tc\gg\Ts$ so that each configuration is independent. Under this assumption, we have
\begin{align}
\Gamma(\tf)\approx N\times\frac{2m^2v^2}{5\pi^2\Mpl^2}G\left(\uv\Ts\right)\,,\label{rate6}
\end{align}
with $\tf=N\Ts+(N-1)\Tc$. If we compared with Eq.~\eqref{rate5}, we see an enhancement factor $N$ come in. Thus, if we take a sufficiently large $N$, decoherence occurs ($\Gamma(\tf)\gg1$) even when the momentum of the massive particle is smaller than the Planck mass, $mv\lesssim\Mpl$. This system might be possible to get carried out in a tabletop experiment with upcoming technology and would open up the possibility of the first detection of the graviton-induced decoherence. 
Thus, we expect that the decoherence caused by the noise of gravitons would offer new vistas to test quantum gravity in tabletop experiments. 

We note that the decoherence rate is enhanced for squeezed states. As we showed in Section~\ref{section3.2}, gravitons produced during inflation experience large squeezing.
Hence, it would be interesting to consider the decoherence induced by 
the primordial gravitons.
We should also notice that our analysis of the superposition state can only be applicable for a frequency range lower than $\xi^{-1}$.
Hence, our estimation of the decoherence rate is a lower bound. 
We should assess the decoherence rate by taking into account the contribution
from a frequency range $\Omega_{\rm m}\xi>1$.

\
\section{Conclusion}
\label{section5}

In this paper, we studied the physical effects induced by gravitons on the system of two massive particles.  
We first derived the effective equation of motion for the geodesic deviation between two particles, that is, the Langevin-type equation in~\eqref{langevin}. We found that gravitons give rise to the noise to the dynamics of particles. Note that  Eq.~\eqref{langevin} without the last term agrees with the one in \cite{Parikh:2020nrd}. We calculated the noise correlation in squeezed coherent states and demonstrated that the squeezed states enhance it compared with the vacuum state. We then discussed the detectability of the noise of gravitons by gravitational wave detectors. It turned out that the amplitude of the noise of gravitons in the case of the Minkowski vacuum is too small to be detected by the current detectors. However, in the squeezed state, the noise of gravitons is enhanced by the exponential factor of squeezing parameter. This is consistent with the amplitude of the stochastic gravitational waves from inflation that experience the large squeezing.

We then explored the effect of gravitons on the process of decoherence in laboratory. We considered a system of two massive particles, one of which is in a superposition state of two spatially-separated locations as shown in Figure~\ref{fig:model} and investigated the loss of coherence of the superposition state  caused by gravitons.  In order to estimate the decoherence rate, we presented two simple configurations of superposition states. In the simplest configuration given in Eq.~\eqref{eq:model}, we found that the decoherence does not occur unless we consider the momentum of massive particle is much larger than the Planck mass ($c=\hbar=1$). As the second simplest configuration, we considered the system that repeats the first configuration Eq.~\eqref{eq:model} $N$ times as time evolves as shown in Figure~\ref{fig:model2}. We found that the decoherence rate is enhanced by the factor $N$ and decoherence happens even when the momentum of massive particle is smaller than the Planck mass if we take a sufficiently large $N$. The system we considered might be possible to get carried out in a tabletop experiment with upcoming technology and would open up the possibility of the first detection of the graviton-induced decoherence. We expect that the graviton-induced decoherence would offer new vistas to test quantum gravity in tabletop experiments.

In order to perform a tabletop experiment to find gravitons, we need a more concrete setup for the experiment.
Correspondingly, we should clarify to what extent other sources of decoherence
contribute and also refine the method for analyzing the decoherence
induced by gravitons.
 We leave these issues for future work.

\section*{Acknowledgments}
S.\,K. was supported by the Japan Society for the Promotion of Science (JSPS) KAKENHI Grant Number JP18H05862.
J.\,S. was in part supported by JSPS KAKENHI Grant Numbers JP17H02894, JP17K18778.
J.\,T. was supported by  JSPS Postdoctoral Fellowship No. 202000912.

\appendix

\section{Momentum integral}
\label{appA}

Here, we calculate the integral in Eq.~(\ref{eom:xi}):
\begin{eqnarray}
\ddot{\xi^i}(t)&=& \frac{1}{2} \ddot{h}^{\rm cl}_{ij}  (0 ,t) \hat{\xi}^j (t)
 -\frac{\kappa}{\sqrt{V}} \hat \xi^j (t) \sum_A\sum_{{\bf k}\leq\Omega_{\rm m}} k^2  e^{A}_{ij} \delta\hat{h}_{\rm I}^A({\bf k},t)\nonumber\\
&&-  \frac{\kappa^2 m}{4}\frac{1}{(2\pi)^3} \hat\xi^j (t) \int^{\Omega_{\rm m}} d^3k\left[ P_{ik} P_{j\ell} +  P_{i\ell} P_{jk} - P_{ij}  P_{k\ell}      \right]   \int_0^t dt'   k\sin k(t-t')  \frac{d^2}{dt^{\prime 2}} \left\{\hat{\xi}^k (t') \hat{\xi}^\ell (t') \right\}   \nonumber \\
&& +\frac{\kappa^2 m}{4}\frac{1}{(2\pi)^3} \int^{\Omega_{\rm m}} d^3k\left[ P_{ik} P_{j\ell} +  P_{i\ell} P_{jk} - P_{ij}  P_{k\ell}      \right] \hat{\xi}^j (t) \frac{d^2}{dt^{ 2}} \left\{\hat{\xi}^k (t) \hat{\xi}^\ell (t) \right\}  \ ,
\label{a1}
\end{eqnarray}
By using the following angular integrals,
\begin{eqnarray}
\int d\Omega=4\pi\,,\qquad\int d\Omega\,k^ik^j=\frac{4}{3}\pi\delta^{ij}\,,\qquad
\int d\Omega\,k^ik^jk^kk^\ell=\frac{4\pi}{15}\left(\delta^{ij}\delta^{k\ell}+\delta^{ik\delta^{j\ell}+\delta^{i\ell}\delta^{jk}}\right)\,,
\end{eqnarray}
we find
\begin{eqnarray}
\int d\Omega \left[ P_{ik} P_{j\ell} +  P_{i\ell} P_{jk} - P_{ij}  P_{k\ell}      \right]
= \frac{8\pi}{5} \left( \delta_{ik} \delta_{j\ell} +  \delta_{i\ell} \delta_{jk} 
                 - \frac{2}{3}  \delta_{ij}  \delta_{k\ell}      \right) \,.
\end{eqnarray}
Then we can perform the momentum integral as
\begin{eqnarray}
\int^{\Omega_{\rm m}} d^3 k \left[ P_{ik} P_{j\ell} +  P_{i\ell} P_{jk} - P_{ij}  P_{k\ell}      \right]
&=&\int_0^{\Omega_{\rm m}} dk\,k^2 \int d\Omega  
 \left[ P_{ik} P_{j\ell} +  P_{i\ell} P_{jk} - P_{ij}  P_{k\ell} \right]\nonumber\\
&=& \frac{8\pi}{15} \left( \delta_{ik} \delta_{j\ell} +  \delta_{i\ell} \delta_{jk} 
     - \frac{2}{3}  \delta_{ij}  \delta_{k\ell}  \right)\Omega_{\rm m}^3 \ .
\end{eqnarray}
If we define the limit representation of Dirac delta function as 
\begin{eqnarray}
f(t-t') =  \frac{\sin  \Omega_{\rm m} (t-t')}{t-t'}\xrightarrow{\Omega_{\rm m}\rightarrow\infty}\pi\delta(t-t')\,,
\end{eqnarray}
we have
\begin{eqnarray}
f(t-t') \big|_{t'=t}&=&\Omega_{\rm m}  \ ,\nonumber\\ 
f(t-t') \big|_{t'=0}&=&\frac{\sin  \Omega_{\rm m} t}{t} \simeq 0 \quad{\rm for}\quad t\neq 0 \,,\nonumber\\
\frac{d}{dt^{\prime }}f(t-t') \big|_{t'=t}&=& 0 \,,\nonumber\\
\frac{d}{dt^{\prime }}f(t-t') \big|_{t'=0}
&=& \frac{\sin  \Omega_{\rm m} t - \Omega_{\rm m} t \cos  \Omega_{\rm m} t}{t^2}\simeq 0 \quad{\rm as}\quad
\Omega_{\rm m}\rightarrow\infty
\,,\label{a6}\\
 \frac{d^2}{dt^{\prime 2}}f(t-t') \big|_{t'=t}&=& -\frac{1}{3} \Omega_{\rm m}^3 \,,\nonumber\\
\frac{d^2}{dt^{\prime 2}}f(t-t') \big|_{t'=0}
&=& \frac{2\sin  \Omega_{\rm m} t - 2\Omega_{\rm m} t \cos  \Omega_{\rm m} t
-\Omega_{\rm m}^2 t^2 \sin  \Omega_{\rm m} t}{t^2}\simeq 0 \quad{\rm as}\quad
\Omega_{\rm m}\rightarrow\infty\,.\quad
\label{a7}
\end{eqnarray}
For Eqs.~(\ref{a6}) and (\ref{a7}), after smearing the functions with an appropriate window function,
these quantities vanish.
The momentum integral  in the second line of the right hand side of Eq.~(\ref{a1}) is written by the function $f(t-t')$ of this form
\begin{eqnarray}
\int_0^{\Omega_{\rm m}} dk k^2  \int_0^t dt'   k\sin k(t-t')  \frac{d^2}{dt^{\prime 2}} \left\{\xi^i (t) \xi^j (t) \right\}   
=   \int_0^t dt'  \frac{d^3}{dt^{\prime 3}}f(t-t')  \frac{d^2}{dt^{\prime 2}} \left\{\xi^i (t') \xi^j (t') \right\}  \ ,
\end{eqnarray}
Using above results, we can evaluate as follows
\begin{eqnarray}
&& \int_0^t dt'  \frac{d^3}{dt^{\prime 3}}f(t-t')  \frac{d^2}{dt^{\prime 2}} \left\{\xi^i (t') \xi^j (t') \right\} \nonumber \\
 && \quad =  \left[  \frac{d^2}{dt^{\prime 2}}f(t-t')  \frac{d^2}{dt^{\prime 2}} \left\{\xi^i (t') \xi^j (t') \right\}\right]_0^t
 -\left[  \frac{d}{dt^{\prime }}f(t-t')  \frac{d^3}{dt^{\prime 3}} \left\{\xi^i (t') \xi^j (t') \right\}\right]_0^t   \nonumber \\
&& \qquad +\left[ f(t-t')  \frac{d^4}{dt^{\prime 4}} \left\{\xi^i (t') \xi^j (t') \right\}\right]_0^t
 - \int_0^t dt'  f(t-t')  \frac{d^5}{dt^{\prime 5}} \left\{\xi^i (t') \xi^j (t') \right\} \nonumber\\
 && \quad = -\frac{1}{3}\Omega_{\rm m}^3 \frac{d^2}{dt^{\prime 2}}\left\{\xi^i (t') \xi^j (t') \right\}+\Omega_{\rm m}^3 \frac{d^4}{dt^{\prime 4}}\left\{\xi^i (t') \xi^j (t') \right\}
 -\frac{\pi}{2}\frac{d^5}{dt^{\prime 5}}\left\{\xi^i (t') \xi^j (t') \right\}\,.
\end{eqnarray}
Then Eq.~(\ref{a1}) becomes
\begin{eqnarray}
\ddot{\xi^i}(t)&=& \frac{1}{2} \ddot{h}^{\rm cl}_{ij}  (0 ,t) \hat{\xi}^j (t)
 -\frac{\kappa}{\sqrt{V}} \hat \xi^j (t) \sum_A\sum_{{\bf k}\leq\Omega_{\rm m}} k^2  e^{A}_{ij} \delta\hat{h}_{\rm I}^A({\bf k},t)\\
&&+\frac{\kappa^2 m}{4}\frac{1}{5\pi^2} \xi^j (t)\left( \delta_{ik} \delta_{j\ell} +  \delta_{i\ell} \delta_{jk}  - \frac{2}{3}  \delta_{ij}  \delta_{k\ell} \right)\left(\Omega_{\rm m} \frac{d^4}{dt^{\prime 4}}\left\{\xi^i (t') \xi^j (t') \right\}
 -\frac{\pi}{2}\frac{d^5}{dt^{\prime 5}}\left\{\xi^i (t') \xi^j (t') \right\}\right).\nonumber
\end{eqnarray}
This is the Eq.~(\ref{langevin}).

\section{Influence functional method}\label{app:infl}
In this appendix, we briefly explain the influence functional method, particularly focusing on its application to the current context. In Appendix~\ref{app:QED}, we consider the QED setup which is the same as the setup discussed in \cite{Breuer and Petruccione}, and obtained the same result. We then generalize the discussion to the gravitational setup in Appendix~\ref{app:gr} to derive Eq.~\eqref{rate}.

\subsection{Influence functional in QED}\label{app:QED}
We consider the superposition state of an electrically charged  particle which is analogous to the setup we discuss in Section~\ref{section4}, and briefly explain how we can compute the coherence of the superposition state after integrating out photons. The setup and the discussion in this appendix is essentially based on \cite{Breuer and Petruccione}.

Specifically, we consider the superposition state of an electrically charged particle such as an electron, and the respective world lines of the particle are parameterized by $(t,{\bf x}_1(t))$ and $(t,{ \bf x}_2(t))$. Corresponding 4-velocities are $v^\mu_a(t)\equiv(1,\dot{\x}_a(t))/\sqrt{1-\dot{\x}_a^2}$ for $a=1,2$, which are normalized as $\eta_{\mu\nu}v_a^\mu v^\nu_a=-1$.

\begin{figure}[tbp]
 \centering
  \includegraphics[width=.5\textwidth, trim=150 80 210 150,clip]{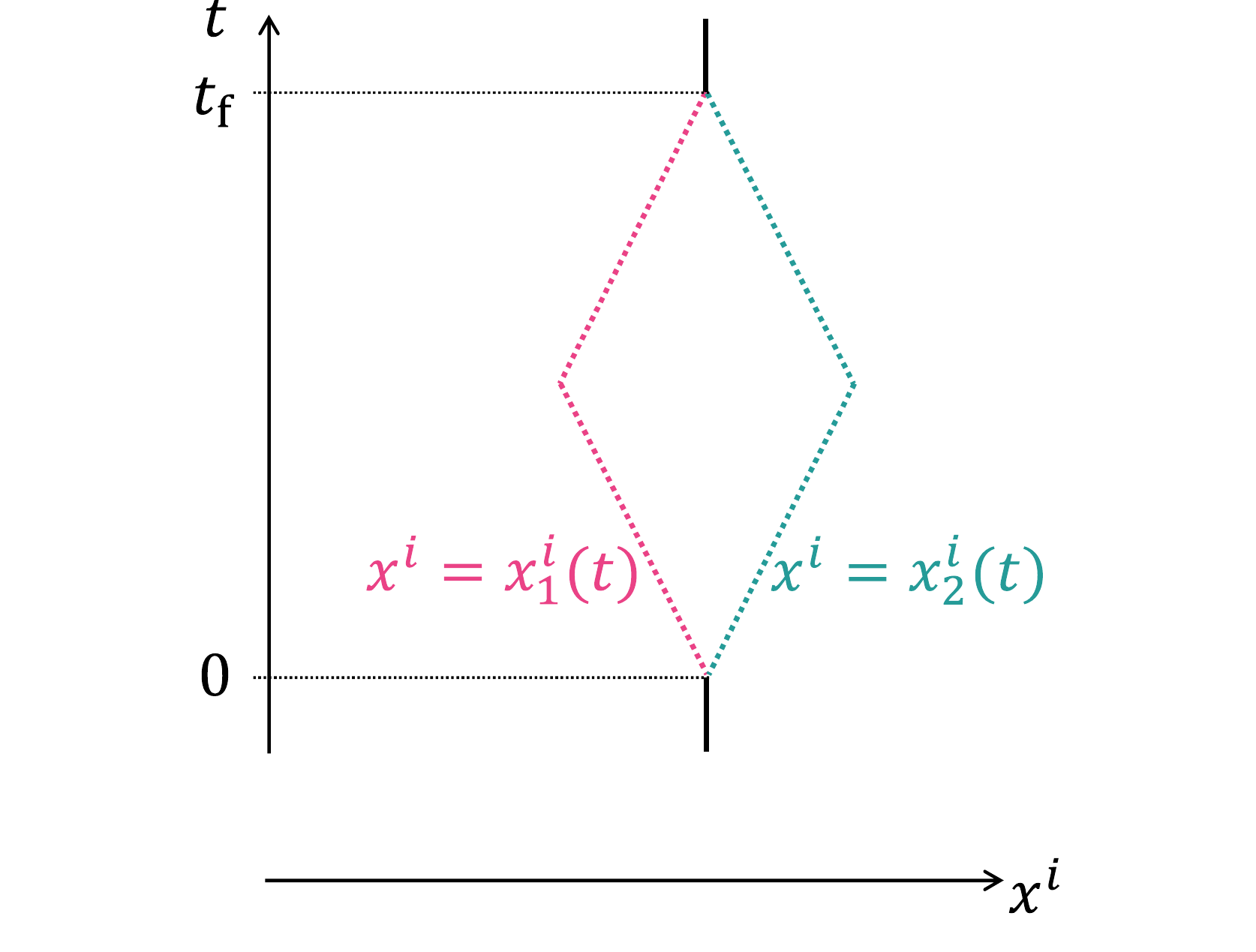}
 \caption{The schematic diagram of each trajectory of a charged particle. The superposition state survives for $0<t<\tf$.}
 \label{fig:QEDmodel} 
\end{figure}

Suppose that the superposition state survives for $0<t<t_{\rm f}$, and the system is in a certain initial quantum state $\ket{\Psi(t_0)}$ at the initial time $t_0<0$. Then, we have $\x_1(t)=\x_2(t)$ for all $t\notin(0,\tf)$. The schematic picture of each trajectory is shown in Figure~\ref{fig:QEDmodel}. We impose the normalization condition $\left<\Psi(t_0)\right|\left.\Psi(t_0)\right>=1$. In this setup, the time evolution of the quantum state in the Schr\"{o}dinger picture can be written as
\begin{align}
&\ket{\Psi(t_0)}\rightarrow\ket{\Psi(t)}=\ket{\Psi_1(t)}+\ket{\Psi_2(t)}\,,\label{eq:evolve1}
\end{align}
where we assume that $\ket{\Psi_1(t)}$ and $\ket{\Psi_2(t)}$ are approximately the eigenstates of the current density operator $\hat J^\mu(\x,t)$. We refer to the respective eigenvalues as $J^\mu_1(\x,t)=(\rho_1(\x,t), {\bf J}_1({\bf x},t))$ and $J^\mu_2(\x,t)=(\rho_2(\x,t), {\bf J}_2({\bf x},t))$, and we treat an electric current $J^\mu$ as an external current. Writing the charge of the particle as $q$, we can write these current densities as $J^\mu_a(\x,t)=qv_a^\mu(t)\delta(\x-\x_a(t))$. We then have $J^\mu_1(\x,t)\neq J^\mu_2(\x,t)$ only for $0< t< \tf$.

These two states are proportional to the initial state $\ket{\Psi(t_0)}$ at the initial time $t_0<0$, and hence we can write 
\begin{align}
\ket{\Psi_a(t_0)}=c_a\ket{\Psi(t_0)}\,,\quad c_1+c_2=1\,,
\end{align}
for $a=1,2$ without loss of generality. Here, $c_a$ can be complex in general. Obviously, this setup is quite similar to the gravitational setup discussed in Section~\ref{section4}. 

\begin{figure}[tbp]
 \centering
  \includegraphics[width=.5\textwidth, trim=120 360 320 150,clip]{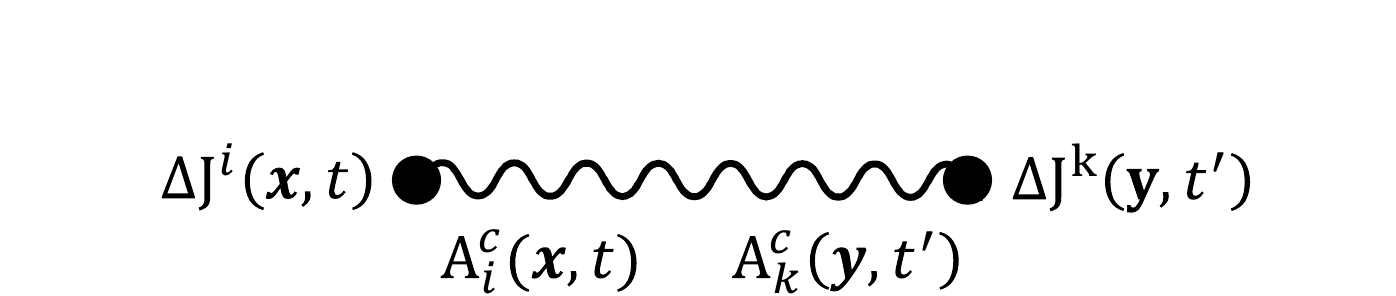}
 \caption{The diagram which contributes to $\Gamma$. Two blobs are the external sources which are given by the difference of external current densities. External sources are connected by the wavy internal line which is the symmetric propagator of photons. } 
 \label{fig:diag1} 
\end{figure}

We now discuss how photons affect the coherence between these states. In the Coulomb gauge which is specified by the condition $\partial_iA^i(\x,t)=0$, the time evolution of the system is governed by the action $S=S[\A,\J]+S_{\rm coul}[\rho]$ which is defined by
\begin{align}
&S[\A,\J]\equiv\int\mathrm{d}^4x\left[-\frac{1}{2}\left(\partial_\mu\A(x)\right)^2+\A(x)\cdot\J(x)\right]\,,\quad S_{\rm coul}[\rho]\equiv-\frac{1}{2}\int\mathrm{d}^4x\int\mathrm{d}^3y\,\frac{\rho(x)\rho({\bf y},t)}{4\pi|\x-{\bf y}|}\,,\label{eq:QEDlag1}
\end{align}
where $x=(\x,t)$. The term $\int\mathrm{d}^4x\,(\A\cdot\J)$ encodes the influence of the external current $\J(x)$ on the time evolution. $S_{\rm coul}$ expresses the influence of the Coulomb energy on the system. Using the standard path-integral expression of the unitary time evolution in the Lagrangian form, one has
\begin{align}
\ket{\Psi_a(t_{\rm f})}=c_a\,e^{iS_{\rm coul}[\rho_a]}\int\mathcal{D}\A\,\left|\A(\x,t_{\rm f})\right>e^{iS[\A,\J_a]}\Psi_0[\A]\,,\quad {\rm for}\quad a=1,2\,. \label{eq:timeev1}
\end{align}
Here, $\Psi_0[\A]\equiv\prod_{\x}\left<\A(\x,0)\right|\left.\Psi(t_0)\right>$ is the initial wave functional. $\left|\A(\x,t)\right>$ denotes an eigenstate of the gauge field $\hat \A(\x)$ in the Schr\"{o}dinger picture: $\hat\A(\x)\left|\A(\x,t)\right>=\A(\x,t)\left|\A(\x,t)\right>$.\footnote{Strictly speaking, the state $\left|\A(\x,t)\right>$ is also an eigenstate of $\hat \J$, while we have omitted its eigenvalue dependence. We can safely omit this dependence here because both of $\ket{\Psi_1(t_{\rm f})}$ and $\ket{\Psi_2(t_{\rm f})}$ are eigenstates of $\hat \J$ with an identical eigenvalue $\J_1(t_{\rm f})=\J_2(t_{\rm f})$.} Using Eq.~\eqref{eq:timeev1}, we obtain the path-integral expression of the decoherence functional as
\begin{align}
&\exp[-\Gamma(t_{\rm f})]=\left|\frac{\left<\Psi_2(t_{\rm f})\right|\left.\Psi_1(t_{\rm f})\right>}{\left<\Psi_2(t_0)\right|\left.\Psi_1(t_0)\right>}\right|\nonumber\\
&=\left|\int\mathcal{D}\A_+\,\int\mathcal{D}\A_-\prod_{\x}\delta\left(\A_+(\x,t_{\rm f})-\A_-(\x,t_{\rm f})\right)e^{i\left(S[\A_+,\J_1]-S[\A_-,\J_2]\right)}\Psi_0[\A_+]\Psi^*_0[\A_-]\right|\,.\label{eq:influence1}
\end{align}
This is nothing but the modulus of the generating functional for photons in the Schwinger-Keldysh formalism. Then, the r.h.s.~of \eqref{eq:influence1} can be obtained by computing all the connected diagrams in which external currents are connected by photon propagators just as we usually do in the external field method. The diagram which contributes to $\Gamma$ is shown  in Figure.~\ref{fig:diag1}, and the result is
\begin{align}
\Gamma(t_{\rm f})=\frac{1}{2}\int^{t_{\rm f}}_0\mathrm{d}t\int^{t_{\rm f}}_0\mathrm{d}t'\int\mathrm{d}^3x\int\mathrm{d}^3y\,\Delta J^i(\x,t)\left<\left\{\hat A_i(\x,t),\hat A_k({\bf y},t')\right\}\right>\Delta J^k({\bf y},t')\,,\label{eq:influence2}
\end{align}
with $\Delta \J\equiv \J_1-\J_2$ and $\hat \A(\x,t)\equiv \hat U^{\dagger}(t,t_0)\hat\A(\x)\hat U(t,t_0)$, $\hat U(t,t_0)$ is a unitary operator expressing the time evolution in the absence of the external current. Here, we used $\Delta J(\x,t)=0$ for all $t\notin(0,t_{\rm f})$, $\left<\A\right>=0$, and the following equality
\begin{align}
\left<\left\{\hat A_i(\x,t),\hat A_k({\bf y},t')\right\}\right>&=\int\mathcal{D}\A_+\,\int\mathcal{D}\A_-\prod_{\x}\delta\left(\A_+(\x,t_{\rm f})-\A_-(\x,t_{\rm f})\right)e^{i\left(S[\A_+,{\bf 0}]-S[\A_-,{\bf 0}]\right)}\nonumber\\
&\hspace{5cm}\times\Psi_0[\A_+]\Psi^*_0[\A_-]A^c_i(\x,t)A^c_k({\bf y},t')\,,
\end{align}
where $\A^c\equiv \frac{\A_++\A_-}{2}$. Eq.~\eqref{eq:influence2} precisely coincides with the one obtained in \cite{Breuer and Petruccione}. Note that when $\left<\A\right>\neq0$, we need to replace $\left<\left\{\hat A_i(\x,t),\hat A_k({\bf y},t')\right\}\right>$ by its connected piece which equals to the symmetric two-point function of $\delta\hat\A\equiv\hat\A-\left<\hat\A\right>$:
\begin{align}
\left<\left\{\hat A_i(\x,t),\hat A_k({\bf y},t')\right\}\right>_{\rm connected}=\left<\left\{\delta\hat A_i(\x,t),\delta\hat A_k({\bf y},t')\right\}\right>\,.
\end{align}

\subsection{Influence functional in gravity}\label{app:gr}
We consider the system discussed in Section~\ref{section4}. 
It is almost obvious that our gravitational setup is formally almost the same as the QED case we discussed in Appendix~\ref{app:QED} except the form of the action. In the gravitational setup, the time evolution of this system is governed by the action \eqref{action-total} in the Fermi normal coordinates, with treating $\xi^i$ as an external field whose time-dependence is given by hand. The last term of the r.h.s.~\eqref{action-total} is the interaction between gravitons and the system consisting of two test particles, which could induce the loss of coherence after integrating out gravitons. The Hamiltonian of the system is still quadratic in the conjugate momentum of gravitons, and hence the derivation of the path-integral expression of the decoherence functional is parallel to the discussion in Appendix.~\ref{app:QED}, leading to
\begin{align}
&\exp[-\Gamma(t_{\rm f})]=\left|\frac{\left<\xi_2(t_{\rm f})\right|\left.\xi_1(t_{\rm f})\right>}{\left<\xi_2(t_0)\right|\left.\xi_1(t_0)\right>}\right|\nonumber\\
&\approx\left|\int\mathcal{D}h^+_{ij}\,\int\mathcal{D}h^-_{ij}\prod_{\x}\delta\left(h^+_{ij}(\x,t_{\rm f})-h^-_{ij}(\x,t_{\rm f})\right)e^{i\left(\tilde S\left[h^+_{ij},\xi^i_1\right]-\tilde S\left[h^-_{ij},\xi^i_2\right]\right)}\Psi_0\left[h^+_{ij}\right]\Psi^*_0\left[h^-_{ij}\right]\right|\,,\label{eq:influence3}
\end{align}
where $\tilde S[h,\xi]$ is 
\begin{align}
\tilde S[h,\xi]=\int dt \int\frac{\mathrm{d}^3k}{{(2\pi)^3}}\sum_{A}  \biggl[ &\frac{1}{2} \dot{h}^{A} ({\bf{k}},t) \dot{h}^{*A}({\bf{k}},t)
 -   \frac{1}{2} k^2 h^A({\bf{k}},t) h^{*A}({\bf{k}},t) \nonumber\\
&+ \frac{m\kappa}{2}\ddot{h}^A ({\bf{k}},t ) \,e_{ij}^A({\bf k}) \xi^i(t) \xi^j(t) \Theta\left(\Omega_{\rm m}-k\right)\biggr] \ . \label{eq:hamiac1}
\end{align}
Here, we took the infinite volume limit $V\to\infty$. We also omitted the terms which only change the overall phase of $\ket{\xi_1(t)}$ and $\ket{\xi_2(t)}$ so that they do not contribute to $\Gamma(\tf)$.  Eq.~\eqref{eq:influence3} implies that we can compute $\Gamma(\tf)$ by writing down all the connected diagrams as in the QED case discussed in Appendx.~\ref{app:QED}. 
The result can be written in terms of the noise of graviton $\delta \hat N_{ij}$ as
\begin{align}
\Gamma (t_{\rm f}) \approx\frac{m^2}{8}\int^{t_{\rm f}}_0\mathrm{d}t\,\Delta(\xi^i\xi^j)(t)\int^{t_{\rm f}}_0\mathrm{d}t'\,\Delta(\xi^k\xi^\ell)(t')\Big\langle\left\{\delta{\hat{N}}_{ij}(t),\,\delta{\hat{N}}_{k\ell}(t')\right\}\Big\rangle\,.
\end{align}
Here, $\Delta(\xi^i\xi^j)(t)\equiv\xi_1^i(t)\xi_1^j(t)-\xi_2^i(t)\xi_2^j(t)$ and $\delta\hat N_{ij}$ is defined by \eqref{eq:noise}. In this way, we obtain Eq.~\eqref{rate}.

\end{document}